\documentclass[prd,twocolumn,showpacs,preprintnumbers,nofootinbib,amsmath,eqsecnum]{revtex4}
\usepackage{graphicx}
\usepackage[section]{placeins}

 \def\gsim{\mathrel{\rlap{\lower4pt\hbox{\hskip1pt$\sim$}}
 \raise1pt\hbox{$>$}}}

 \newcommand\la{\langle}
 \newcommand\ra{\rangle}
 \newcommand\beq{\begin{equation}}
 
 \newcommand\eeq{\end{equation}}
 \newcommand\beqn{\begin{eqnarray}}
 \newcommand\eeqn{\end{eqnarray}}
 \newcommand\eq{\!=\!}
 \newcommand\eqq{\!\equiv\!}
\def\mb{\,\mbox{mb}}
\def\fm{\,\mbox{fm}}
\def\GeV{\,\mbox{GeV}}
\def\TeV{\,\mbox{TeV}}
\def\lsim{\mathrel{\rlap{\lower4pt\hbox{\hskip1pt$\sim$}}
    \raise1pt\hbox{$<$}}}         
\def\gsim{\mathrel{\rlap{\lower4pt\hbox{\hskip1pt$\sim$}}
    \raise1pt\hbox{$>$}}}         

\def\mb{\,\mbox{mb}}
\def\fm{\,\mbox{fm}}
\def\GeV{\,\mbox{GeV}}

\def\beq{\begin{equation}}
\def\eeq{\end{equation}}

\def\beqy{\begin{eqnarray}}
\def\eeqy{\end{eqnarray}}

\begin{document}

\preprint{LU-TP 14-27}

\title{\bf A heuristic description of high-\boldmath$p_T$ hadron
production in heavy ion collisions}

\author{Jan Nemchik$^{1,2}$}
\author{Roman Pasechnik$^{3}$}
\author{Irina Potashnikova$^4$}

\affiliation{{$^{1}$\sl 
Czech Technical University in Prague,
FNSPE, B\v rehov\'a 7,
11519 Prague, Czech Republic
}\\
{$^{2}$\sl
Institute of Experimental Physics SAS, Watsonova 47,
04001 Ko\v sice, Slovakia
}\\
{$^{3}$\sl
Department of Astronomy and Theoretical Physics, Lund
University, SE-223 62 Lund, Sweden
}\\
{$^4$\sl Departamento de F\'{\i}sica,
Universidad T\'ecnica Federico Santa Mar\'{\i}a;\\
Centro Cient\'ifico-Tecnol\'ogico de Valpara\'{\i}so,
Casilla 110-V, Valpara\'{\i}so, Chile}
}
%
%
%
%
%
\begin{abstract}
Using a simplified model for in-medium dipole evolution accounting for 
color filtering effects we study production of hadrons at large 
transverse momenta $p_T$ in heavy ion collisions. In the framework of 
this model, several important sources of the nuclear suppression 
observed recently at RHIC and LHC have been analysed.
A short production length of the leading hadron $l_p$ causes
a strong onset of color transparency effects manifested themselves 
as a steep rise of the nuclear modification factor $R_{AA}(p_T)$ at 
large hadron $p_T$'s. A dominance of quarks with higher $l_p$ leads 
to a weaker suppression at RHIC than the one observed at LHC.
In the RHIC kinematic region we include an additional suppression 
factor steeply falling with $p_T$, which is tightly related to the energy conservation 
constraints. The latter is irrelevant at LHC up to $p_T\lsim 70\,\GeV$ 
while it causes a rather flat $p_T$ dependence of the $R_{AA}(p_T)$ factor 
at RHIC c.m. energy $\sqrt{s} = 200\,\GeV$ and even an increasing 
suppression with $p_T$ at $\sqrt{s} = 62\,\GeV$. The calculations 
contain only a medium density adjustment, and for an initial time scale 
$t_0$ = 1$\fm$ we found the energy-dependent maximal values of 
the transport coefficient, $\hat{q}_0 = 0.7, 1.0$ and $1.3\,\GeV^2/\fm$
corresponding to $\sqrt{s} = 62, 200\,\GeV$ and $2.76\,\TeV$, 
respectively. We present a broad variety of predictions for the nuclear modification
factor and the azimuthal asymmetry which are in a good agreement with 
available data from experiments at RHIC and LHC.
\end{abstract}


\maketitle

%
\section{Introduction}
%

The available data on high-$p_T$ hadron production 
in heavy ion collisions at the LHC 
\cite{alice-new,cms-new1,cms-new2} 
and RHIC \cite{phenix-b,phenix-0,phenix62-b,star} 
clearly demonstrate a strong nuclear suppression 
exhibiting rather different features at 
various c.m. energies. The nuclear modification 
factor $R_{AA}$ measured at the LHC reaches significantly 
smaller values than those at RHIC. Simultaneously, the 
$R_{AA}(p_T)$ factor steeply rises with $p_T$ at the LHC 
while it exhibits a rather flat $p_T$ dependence at 
RHIC c.m. energy $\sqrt{s} = 200\,\GeV$ 
and even a significant fall at $\sqrt{s} = 62\,\GeV$.

An explanation of the observed high-$p_T$ suppression 
may be connected to our understanding of the hadronization 
phenomenon \cite{within}, namely, an energy loss scenario 
of a created color parton after a heavy ion collision 
discussed in detail e.g. in Ref.~\cite{green} (see also
Sects.~\ref{hadronization} and \ref{eloss}). 
Production of such a parton with a high transverse momentum 
initiates the hadronization process, 
which is finalized by the formation of a jet 
of hadrons. In this paper, we are focused on 
a specific type of jets when the main 
fraction $z_h$ of the jet energy $E$ is carried 
by a single (leading) hadron. 
Here, $z_h$ of the detected hadron cannot be measured. 
However, the convolution of the jet momentum dependence and the 
hadron fragmentation leads to typically large values $z_h\gsim 0.5$
\cite{green} as is demonstrated in Sect.~\ref{eloss} (see also
Fig.~\ref{fig:mean-zh}). The initial virtuality of the jet is of the same order as 
its energy scale. Thus at large jet energies a significant 
dissipation of energy via gluon radiation takes place. 
The latter presumably happens at an early stage of hadronization
at small space-time separations (see e.g. Ref.~\cite{green}).

There are two time scales controlling the hadronization process
\cite{within,kn-review} as is discussed in Sect.~\ref{hadronization}. 
The first scale is connected to the energy 
conservation in hadron production at large $z_h$. 
The latter scale controls production of a colorless 
dipole configuration at late stages of hadronisation which is called 
sometimes as ``pre-hadron'' or QCD dipole.
The formation of such a ``pre-hadron'' effectively terminates
 the radiative energy dissipation into gluon emissions. 
 
The corresponding length scale $l_p$ called the production length 
was calculated earlier in \cite{jet-lag,green} within 
a perturbative hadronization model. As will be shown 
in Sect.~\ref{eloss} (see also Fig.~\ref{fig:mean-lp}) this length scale appears 
to be short such that the formation of a ``pre-hadron'' may happen in the medium. 
Besides, this length appears to be weakly dependent on dipole $p_T$ 
as a result of two main effects working in the opposite 
directions. Indeed, while the Lorentz factor effectively stretches 
the $l_p$ scale at larger $p_T$, an increase of the energy loss 
rate with $p_T$, in opposite, causes a shortening of $l_p$.

The short $l_p$ scale means that the produced colorless dipole (hadronic state) 
has to survive during its propagation in the medium, otherwise it can not be detected. 
Evolution of such a dipole while it propagates in the medium 
up to the moment of formation of the final-state hadronic wave function 
is controlled by the second time scale called the formation time $t_f$
(or formation length $l_f$) as is discussed in Refs.~\cite{within,green}
and also in Sect.~\ref{hadronization}. Here, the color transparency (CT) 
mainly controls the surviving probability of the propagating dipole since
the medium appears to be more transparent for smaller dipoles \cite{zkl}. 
A relation between the transport coefficient which 
characterizes the medium density and the universal dipole cross
section has been found in Refs.~\cite{broadening,jkt}. So the transport
coefficient can be probed by observable effect of dipole attenuation.
For our previous work on this topic, see Ref.~\cite{green}.

The CT effect in production of high-$p_T$ hadrons in heavy ion collisions 
was calculated within a rigorous quantum-mechanical 
description based on the path-integral formalism in Ref.~\cite{green}.
In order to avoid complicated numerical calculations but to retain 
a simple physical understanding of the main features of underlying 
dynamics, in the present paper we start from a simplified model of 
Ref.~\cite{my-alice}. Further, we generalize this model also 
for non-central heavy ion collisions at various energies 
with different contributions of quark and gluon jets to the leading 
high-$p_T$ hadron production (see Sect.~\ref{dipole}).
In addition, we incorporate the color filtering effects describing 
an expansion of the dipole in a medium during the formation time. 
We found an analytical solution of the corresponding evolution 
equation (\ref{410}) for the mean dipole size. Our manifestly simple 
formulation enables us to study the predicted nuclear modification 
factor $R_{AA}$ with respect to various data as is presented in Sect.~\ref{data}.  
The maximal value of the transport coefficient, $\hat q_0$, corresponding to
central heavy nuclei collisions at a fixed energy is the only free parameter 
of the model and it is universal and independent on a particular observable. 

By a comparison with the data, we found that for initial time $t_0 = 1\,\fm$
the values of this parameter range between $\hat q_0 = 0.7 \GeV^2 \!/ \!\fm$ 
at $\sqrt{s}= 62\GeV$ and $1.3 \GeV^2 \!/ \! \fm$ at $\sqrt{s}= 2.76\TeV$.
These values of $\hat q_0$ are larger than the value found
in Ref.~\cite{my-alice} due to color filtering effects in
dipole size evolution, Eq.~(\ref{410}). Simultaneously,
they are only slightly smaller than those in Ref.~\cite{green} 
within a rigorous quantum-mechanical description.
However, all the values of $\hat{q}_0$ found within both 
the simplified model \cite{my-alice} and the
rigorous quantum-mechanical description \cite{green}
are smaller by an order of magnitude than in Ref.~\cite{phenix-theor} where
the prediction was based on the pure energy loss 
approach \cite{miklos} due to a shorter production length is our approach. 
It is worth emphasizing that our approach, 
based on perturbative QCD (pQCD), is irrelevant to data at small $p_T\lesssim 6\GeV$, 
which are dominated essentially by hydrodynamics. 

At large values of $x_L\eqq x_F\eq2 p_L/\sqrt{s}$ and/or 
$x_T \eq 2p_T/\sqrt{s}$ we incorporate an additional effect 
related to initial state interaction (ISI) energy deficit
as was described in Refs.~\cite{isi,isi-jan,kn-review,green}
and is presented in Sect.~\ref{isi}. The corresponding enhanced nuclear 
suppression is predicted to be important in the $p_T$ dependence 
of $R_{AA}$ factor at RHIC c.m. energies $\sqrt{s}=200\GeV$ and 
$62\GeV$. For the LHC kinematics, this effect causes a levelling of 
the $R_{AA}(p_T)$ behavior at the maximal measured 
$p_T\gsim 70\div 100\,\GeV$. Finally, as a complementary 
test of our approach, in Sect.~\ref{v2} we compared the results on the
azimuthal anisotropy of produced hadrons with the corresponding 
RHIC/LHC data and a good agreement has been found.

%
\section{Remarks about hadronization}
\label{hadronization}
%

The subject of hadronization has been discussed in detail during
last twenty years in numerous papers (see e.g. Refs.~\cite{knp,within,pert-hadr,
jet-lag,eloss,kn-review,green,conservation14}). Here we present the main 
features of nuclear suppression which differ from the popular interpretation 
based on the parton energy loss concept.

Production of single hadrons with high transverse momenta $p_T$ in 
heavy ion collisions at high energies of RHIC \cite{phenix-b,star} and LHC 
\cite{alice-new,cms-new1,cms-new2} clearly demonstrates a strong 
suppression compared to that in $pp$ collisions. While there is a 
consensus about the source of this suppression which is due to 
final-state interactions with the co-moving medium created in 
a collision, the mechanism of such interactions is still under intensive 
debates.
 
The popular interpretation of the observed high-$p_T$ hadron
suppression is based on the loss of energy by a parton propagating 
through the medium created in a collision. The perturbative radiative 
energy loss is caused by the ``wiggling" of the parton 
trajectory due to multiple interactions in the medium. 
Every time when the parton gets a kick from a scattering 
in the medium, a new portion of its color field is shaken off.
The loss of energy induced by multiple interactions is naturally related 
to the broadening of the parton transverse (relative to its trajectory, 
i.e. to $\vec p_T$) momentum $k_T$ \cite{bdmps},
\beq
\frac{dE}{dL}=-\frac{3\alpha_s}{4}\,\Delta k_T^2(L)=
-\frac{3\alpha_s}{4}\int\limits_0^L dl\, \hat q(l)\,,
\label{100} 
\eeq
where $\hat q(l)$ is the rate of broadening $\Delta k_T^2$, 
which may vary with $l$ along the parton trajectory,
\beq
\hat q(l)=\frac{d\Delta k_T^2}{dl}\,.
\label{120}
\eeq

However, a natural expectation within the energy loss scenario, 
that dissipation of energy by the parton in a medium should suppress 
production of leading hadrons, raises several important questions.

In particular, one usually assumes that the energy loss induces 
a shift $\Delta z_h$ in the argument of the fragmentation function, i.e. 
$z_h\Rightarrow z_h+\Delta z_h$. This could be true, if hadronization 
of the parton is initiated outside the medium. However, in practice 
it may start earlier and the main part of gluon radiation occurs 
at short distances right after the hard collision (see Sect.~\ref{eloss} and 
Fig.~\ref{fig:mean-lp}).
 
Another related assumption within the energy loss scenario 
which has never been rigorously justified, is that the path of the parton 
propagating in the medium is assumed to be always longer than 
the medium size. According to this assumption the colorless hadronic 
state, which does not radiate energy any longer and is eventually
detected, should be produced outside the medium. The validity 
of this assumption should further be investigated and the path 
length available for hadronization should be evaluated.
This issue was discussed in Ref.~\cite{within} in a particular case 
of semi-inclusive deep-inelastic scattering (SIDIS), as well as 
within dynamical models of hadronization \cite{pert-hadr,jet-lag} 
providing rather solid constraints on the above assumption.

\begin{figure}[htb]
   \vspace*{-10mm}
   \centerline{
         \scalebox{0.41}{\includegraphics{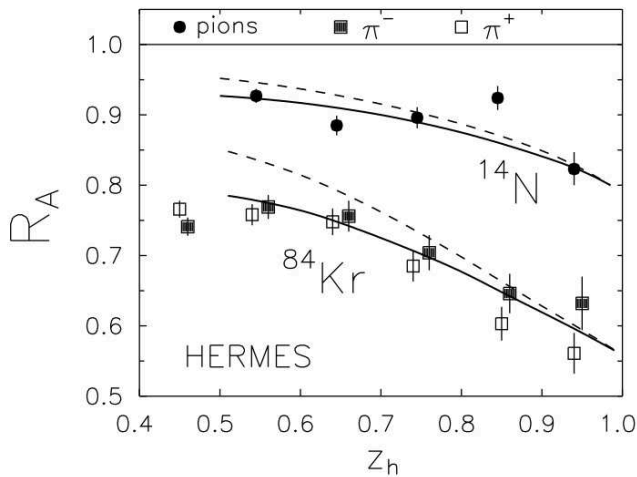}}}
   \caption{\label{fig:mine}
   A comparison of the predicted \cite{knp,within} 
   $z_h$-dependence of the nuclear suppression factor 
   in inclusive electroproduction of pions with the HERMES data
   \cite{hermes1,hermes2}. Solid and dashed curves show 
   the results with and without energy loss 
   corrections, respectively.}
\end{figure}

Besides pure theoretical arguments, the energy loss scenario becomes problematic 
in explaining, for example, production of hadrons in SIDIS. The latter process 
provides a rigorous test for in-medium hadronization models in a more definite 
environment than in heavy ion collisions. Here, the medium density and geometry 
are well known and time independent and the hadron fractional momentum $z_h$ 
(the argument of the fragmentation function) has been measured.
The corresponding predictions in a model including an analysis 
of the hadronization length \cite{knp}, which was found rather short, 
were made five years prior the measurement and then were successfully 
confirmed by the HERMES experiment \cite{hermes1,hermes2}.
The comparison with data shown in Fig.~\ref{fig:mine} 
(see also Ref.~\cite{within}) demonstrates a rather good agreement 
of these predictions and the data.
%
%
%
%
%
\begin{figure}[htb]
   \vspace*{-5mm}
   \centerline{
         \scalebox{0.38}{\includegraphics{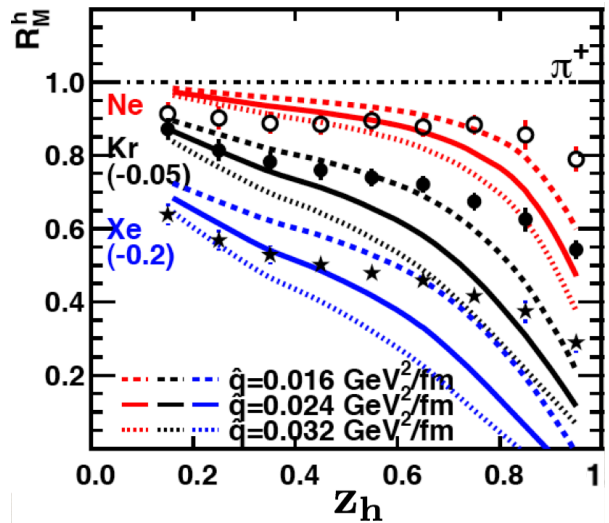}}}
   \caption{\label{fig:wang}
    (color online)
   The results of description of the HERMES data
   by the model \cite{wang-sidis} based on the energy 
   loss scenario, i.e. assuming a long time of hadronization. 
   The curves for several values of the transport coefficient 
   are presented. }
\end{figure}
%

On the other hand, attempts to explain the HERMES
data within the energy loss scenario were not successful. 
An example of comparison of the model \cite{wang-sidis} with 
the data \cite{hermes2} is depicted in Fig.~\ref{fig:wang}.
Such a comparison clearly demonstrates that the model 
fails to explain the data at large $z_h>0.5$ which dominates the
high-$p_T$ hadron production in heavy ion collisions (see Fig.~\ref{fig:mean-zh}). 
Even adjustments of the transport coefficient $\hat q$ (actually
well-known for the cold nuclear matter studies \cite{broadening,domdey}) 
do not improve the situation.

\begin{figure}[htb]   
   \vspace*{35mm}
   \centerline{
         \scalebox{0.32}{\includegraphics{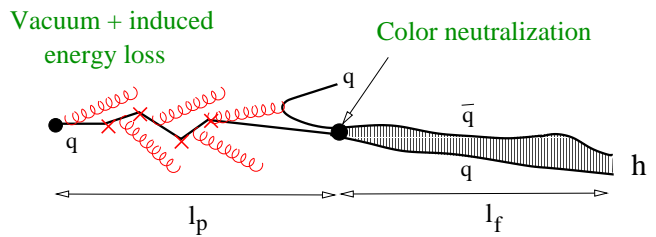}}}
    \caption{\label{fig:space-time}
    (color online)
    Space-time development of hadronization of a highly virtual 
    quark producing a leading hadron carrying a substantial fraction 
    $z_h$ of the initial light-cone momentum.}
\end{figure}

Thus, the assumption of a long hadronization length should be checked more 
carefully and thoroughly. The basic space-time scales of the in-medium hadronization 
process are indicated schematically in Fig.~\ref{fig:space-time}.
%
%
%
The quark regenerating its color field, which has been stripped off 
in a hard scattering, intensively radiates gluons and dissipates its energy, 
either in vacuum or in a medium. Multiple interactions in the medium 
induce an additional, usually less intensive, radiation.
The loss of energy ceases at the moment called the production time $t_p$ 
(or production length $l_p \equiv t_p$) when the quark picks up an 
antiquark neutralizing its color. Thus, the production length $l_p$ is actually 
the distance at which the color neutralization occurs and a colorless dipole 
(also called ``pre-hadron'') is produced and starts developing a wave function.
The produced colorless dipole has yet neither hadron wave function nor hadron 
mass, and it takes the formation time $t_f$ to develop both. So this process 
is characterized by the formation length which usually is rather long, 
$l_f\sim 2E/(m_{h^*}^2-m_h^2)$ \cite{kz91,within}.

The hadronization mechanism covers only the first stage of hadron production. 
In what follows, we concentrate on the production length scale $l_p$, which is 
the only path available for energy loss. The formation stage is described within 
the path-integral method in Ref.~\cite{green} and is treated within the simplified 
model in Sect.~\ref{dipole}.

Notice that the question whether the hadronization occurs 
inside or outside the medium might have no definite answer. 
Indeed, there is a typical quantum-mechanical uncertainty such that
the production amplitudes with different values of $l_p$ interfere.
Such an interference has been studied in Ref.~\cite{avt} for SIDIS 
and a noticable effect has been found. An extension 
of those results for a hot medium is a big challenge, so 
in what follows we disregard the interference and employ instead
the effective standard semi-classical space-time picture of the 
hadronization process.

%
\section{Radiative energy loss in vacuum}
\label{eloss}
%

First of all, one should discriminate between vacuum and 
medium-induced radiative energy loss. High-$p_T$ partons radiate 
gluons and dissipate energy even in vacuum, and the corresponding
rate of energy loss may considerably exceed the medium-induced
value Eq.~(\ref{100}) because the former is caused by a hard collision.

%
\subsection{\boldmath Regeneration of the parton color field}
%

A high-$p_T$ scattering of partons leads to an intensive gluon
radiation in forward-backward directions, in which the
initial color field of the partons is shaken off due to the strong
acceleration caused by the hard  collision. The radiated
gluons accompanying the colliding partons do not survive
the hard interaction and lose their coherence up to transverse momenta
$k_T\lsim p_T$. Therefore, the produced high-$p_T$ parton is lacking
this part of the field and starts regenerating it via radiation of a new cone 
of gluons which are aligned along the new direction.
One can explicitly see the two cones of gluon radiation 
in the Born approximation calculated in Ref.~\cite{gunion-bertsch}.
This process lasts for a long time proportional to the jet energy
($E\approx p_T$). 

Let such a jet be initiated by a quark. Then the coherence length (or time) 
$l_g$ of the gluon radiation depends on the gluon fractional light-cone momentum 
$\beta$ and its transverse momentum $k_T$ relative to the jet axis as,
%
\beq
l_g=\frac{2E}{M_{qg}^2-m_q^2}= \frac{2E\beta(1-\beta)}{k_T^2+x^2\,m_q^2}\, ,
\label{140}
\eeq
%
where $M_{qg}$ is the invariant mass of
the recoil quark and radiated gluon system.
Then using Eq.~(\ref{140}) one can trace how much energy 
is lost over the path length $L$ via gluons which have 
lost their coherence (i.e. were radiated) during this time 
interval,
%
\beq
\frac{\Delta E(L)}{E} =
\int\limits_{\Lambda^2}^{Q^2}
dk_T^2\int\limits_0^1 d\beta\,\beta\,
\frac{dn_g}{d\beta\,dk_T^2}
\Theta(L-l_g)\, ,
\label{160}
 \eeq
%
where $Q^2\sim p_T^2$ is the initial quark virtuality; 
the infra-red cutoff is fixed at $\Lambda=0.2\, \GeV$ and
the radiation spectrum reads
%
\beq
\frac{dn_g}{dx\,dk_T^2} =
\frac{2\alpha_s(k_T^2)}{3\pi\,\beta}\,
\frac{k_T^2[1+(1-\beta)^2]}{[k_T^2+\beta^2 m_q^2]^2}\, .
\label{180}
\eeq
%
%
%
\begin{figure}[htb]   
   \vspace*{-25mm}
   \centerline{
         \scalebox{0.43}{\includegraphics{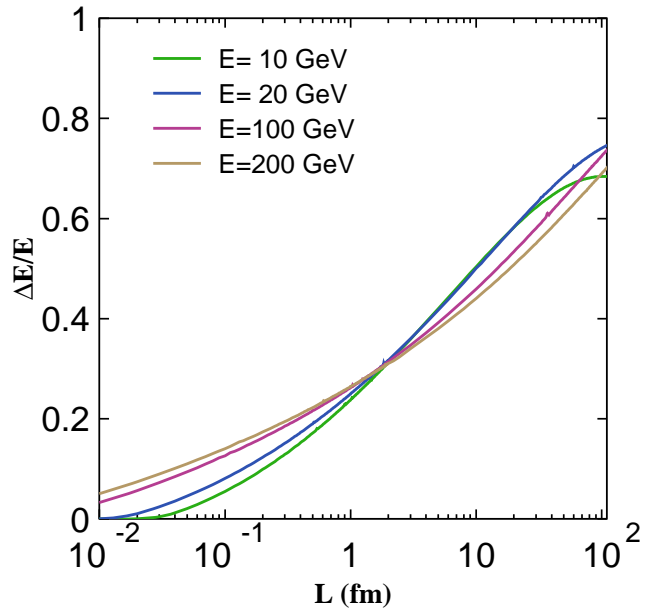}}}
    \caption{\label{fig:e-loss}
    (color online)
    The fractional energy loss by a quark with different initial 
    energies in vacuum vs path length $L$.}
\end{figure}
%

Fig.~\ref{fig:e-loss} shows the fractional vacuum energy loss 
by a quark vs distance from the hard collision
for several initial energies $10,\  20,\ 100,$
and $200\,\GeV$ (compare with heavy flavors in Ref.~\cite{eloss}).
The energy dissipation rate is considerable and energy conservation 
may become an issue for a long path length if one wants to produce 
a leading hadron. Indeed, the production rate of high-$p_T$ hadrons
is determined by a convolution of the parton distributions 
in the colliding hadrons (which suppresses large fractional 
momenta $\beta$ at high $p_T$) with the transverse 
momentum distribution in the hard parton collisions 
(also suppresses large $p_T$) and with the fragmentation 
function $D(z_h)$ of the produced parton.
The latter has maximum at small $z_h\ll1$, which, however, 
is strongly suppressed by the convolution, pushing the maximum 
towards large values of $z_h$. 

\begin{figure}[htb]
   \centerline{
         \scalebox{0.43}{\includegraphics{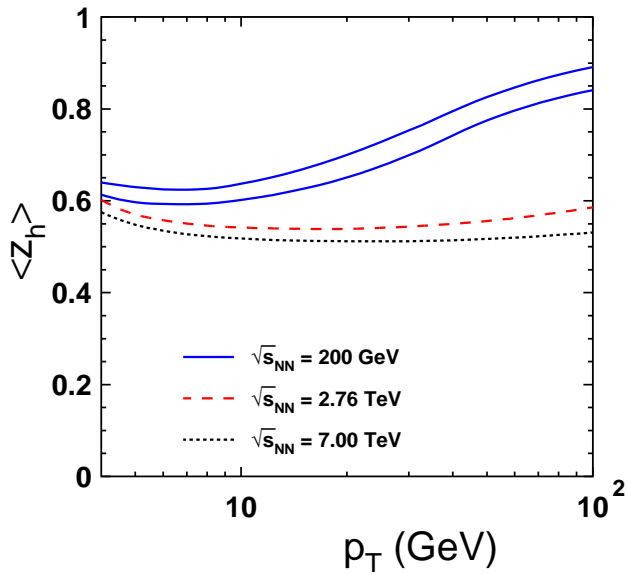}}}
   \caption{\label{fig:mean-zh}
   (color online)
   The mean fraction $\la z_h\ra$  of the jet energy 
   carried by a hadron detected with transverse momentum $p_T$. 
   The calculations are performed for collision energies
   $\sqrt{s}=0.2,\ 2.76$ and $7\TeV$.}
\end{figure}

Numerical results of the convolution for the mean value $\la z_h\ra$ 
\cite{my-alice,green} are depicted in Fig.~\ref{fig:mean-zh}, separately 
for quark and gluon jets (upper and bottom solid curves) and at different 
energies, $\sqrt{s}=200,\ 2760$ and $7000\GeV$.
We see that the lower is the collision energy, the larger is $\la z_h\ra$, 
especially at high $p_T$, since the parton $k_T$ distribution gets steeper. 
For LHC energies the magnitude of $\la z_h\ra$ practically 
saturates as a function of $\sqrt{s}$ and $p_T$, and
becomes indistinguishable for quark and gluonic jets.
%
%
%

We would like to emphasize the difference between 
inclusive production of a high-$p_T$ hadron and 
a high-$p_T$ jet. According to the discussion above,
in the former case the detected hadron carries the main fractional 
light-cone momentum $z_h$ of the parent jet energy.
In the latter case, only the whole jet transverse 
momentum $p_T^{\rm jet}$ is required to be large and no other 
constraints are imposed. Then the fractional momenta of 
hadrons in the jet are typically very low, so energy conservation 
does not cause any severe constraints on the hadronization 
time scale which inceases with $p_T$ and may be rather large.

%
\subsection{How long does it take to produce a hadron?}
%

Production of a hadron with a large fractional momentum 
$z_h$ becomes impossible when the parton radiates 
a substantial fraction of its initial energy $E$, 
$\Delta E/E>1-z_h$. Thus, energy conservation imposes 
an upper bound on the production length $l_p$.
Figures \ref{fig:e-loss} and \ref{fig:mean-zh} 
clearly demonstrate that such a maximal value of $l_p$ 
is rather short and nearly independent of $p_T$. The latter
seems to be in contradiction with the Lorentz factor, 
which should lead to $l_p\propto p_T$. However, the intensity 
of gluon radiation and the energy dissipation rate rise
as well, approximately as $p_T^2$, leading to an 
opposite effect of $l_p$ decrease.
%
%
\begin{figure}[thb]
\vspace*{0.4cm}
   \centerline{
         \scalebox{0.44}{\includegraphics{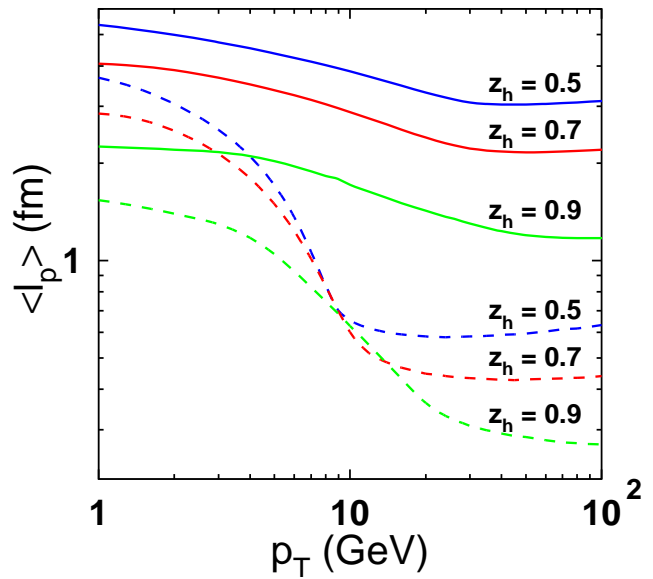}}}
   \caption{\label{fig:mean-lp}
   (color online)
   The mean production length as function of energy 
   for quark (solid curves) and gluon (dashed curves) jets.
   In both cases the curves are calculated at 
   $z_h=0.5,\ 0.7,\ 0.9$ (from top to bottom). }
\end{figure}
%

More precisely, the $p_T$ dependence of $\la l_p\ra$ 
can be derived  within a dynamic model of hadronization. 
It was done in Ref.~\cite{jet-lag,my-alice} employing 
a perturbative treatment of hadronization \cite{pert-hadr}.
Some numerical results are plotted in Fig.~\ref{fig:mean-lp} 
for fragmentation of quarks and gluons by solid and 
dashed curves, respectively. We see that the mean production 
length is rather short and slowly decreases with $p_T$. 
The production length for gluon jets is shorter due to 
a more intensive vacuum energy loss and a stronger 
Sudakov suppression, which leads to a reduction 
of $\la l_p\ra$. 
 
The  production length in Fig.~\ref{fig:mean-lp}  
demonstrates a decreasing trend with $p_T$,
which is in variance with the naive expectation of a rise 
due to the Lorentz factor. As was explained above, 
this happens due to the growing virtuality and radiative 
dissipation of energy.

Note, so far we have taken into consideration only 
the energy dissipation in vacuum. Apparently, by adding the medium 
induced energy loss one can only enhance the energy deficit 
and make the production length even shorter.

%
\section{In-medium dipole evolution: a detailed description}
\label{dipole}
%

The dipole attenuation and evolution is a dense medium has first been studied
in the framework of sophisticated path integral approach in Ref.~\cite{green}.
Here we follow the same notations and develop a much simplified heuristic 
description of the dipole dynamics in the medium additionally incorporating 
the color filtering effects in high-$p_T$ hadron production. The suggested 
approach accounts for all the major physical effects
in a transparent manner while being very simple to use in practice and having a single 
adjustable parameter -- the maximal value of the transport coefficient characterizing 
the medium properties.

In vacuum (e.g. in $pp$ collisions), in evaluations of the invariant cross section 
of inclusive hadron production we use the model of Ref.~\cite{wang}:
\begin{widetext}
\beqn
\frac{d\sigma_{pp}}{dy\,d^2p_T} &=&
K\sum_{i,j,k,l}\,
\int dx_i dx_j d^2k_{iT} d^2k_{jT}
F_{i/p}(x_i,k_{iT},Q^2)\,
F_{j/p}(x_j,k_{jT},Q^2)
\frac{d\sigma}{d{\hat t}}(ij\to kl)\,
\frac{1}{\pi\, z_h}\,D_{h/k}(z_h,Q^2)\,,
\label{300}
\eeqn
\end{widetext}
%
which corresponds to the collinear factorization expression modified by an
intrinsic transverse momentum dependence. In Eq.~(\ref{300}), 
$d \sigma (ij \to kl ) / d{\hat t}$ is the hard scattering cross section, and
we follow the notations in Ref.~\cite{wang}. 
Then the distribution in $k_T$ reads
%
\beq
F_{i/p}(x, k_T, Q^2) = 
F_{i/p}(x, Q^2)\;
g_p(k_T, Q^2)\, ,
\label{320}
\eeq
%
where
%
\beq
g_p(k_T, Q^2) =
\frac{1}{\pi\left \la k_T^2(Q^2)
\right \ra}\;e^{-k_T^2 / \left \la 
k_T^2 (Q^2) \right \ra}\, .
\label{330}
\eeq
%
The dependence of $\left \la k_T^2 (Q^2) \right \ra$ on $Q^2$ scale was suggested 
in Ref.~\cite{wang} as follows $\la k_T^2 \ra_N(Q^2) = 1.2 \GeV^2 + 
0.2 \alpha_s(Q^2)Q^2$. We use the phenomenological parton distribution functions (PDFs) 
$F_{i/p}(x,Q^2)$ from the MSTW08 leading order (LO) fits \cite{mstw}. For the fragmentation 
function $D_{h/k}(z_h,Q^2)$ we rely on the LO parametrization given in Ref.~\cite{florian}.
As was explicitly checked in Ref.~\cite{green}, this model describes well the data 
on $p_T$-dependence of pion production in $pp$ collisions at $\sqrt{s}=200\GeV$ 
and charged hadron production at $\sqrt{s}=7\TeV$.

The produced leading hadron carries a significant fraction $z_h\gsim 0.5$ of the initial 
light-cone parton momentum \cite{green} as was discussed in the previous Sect.~\ref{eloss} 
and presented in Fig.~\ref{fig:mean-zh}. The energy conservation requires that the hadronization 
process should be terminated promptly by the color neutralization, i.e. by production 
of a colorless ``pre-hadron'' (dipole), otherwise the leading parton looses too much of its 
energy such that it becomes unable to produce a hadron with a large $z_h$ fraction any longer. 
The corresponding time scale for production of a colorless dipole is thereby rather short and 
practically does not rise with $p_T$, as was demonstrated in Fig.~\ref{fig:mean-lp}.

In comparison with the vacuum case,  the multiple interactions of the parton in 
a dense medium induce an extra energy loss. The latter reduces the 
production time even more \cite{within}. Thus, this effect enables us to calculate 
the corresponding survival probability function $W$ this is the probability for a ``pre-hadron'' 
to leave the dense medium. This is according to the CT effect \cite{zkl} when 
the attenuation rate of a dipole depends on the dipole separation $r$ quadratically at $r\to 0$, i.e.
%
\beq
\left. \frac{dW}{dl} \right|_{r \to 0}= -
{1\over2}\,\hat q(l)\, r^2\,,
\label{302}
\eeq
%
where the rate of broadening $\hat q(l)$ given by Eq.~(\ref{120}) 
is typically called the transport coefficient which determined properties 
of the medium (see e.g. Ref.~\cite{bdmps}). Here, we rely on the widely used naive model for $\hat q$
which is taken to be dependent on the number of participants 
$n_{part}(\vec b,\vec\tau)$, on impact distance $b$
and the length scale $l$ as follows \cite{frankfurt},
%
\beq 
\hat q(l,\vec b,\vec\tau)=\frac{\hat
q_0\,l_0}{l}\, \frac{n_{part}(\vec b,\vec\tau)}{n_{part}(0,0)}
\,\Theta(l-l_0)\,,
\label{340} 
\eeq
%
where $\vec  b$ and $\vec \tau$ are the impact parameters of nuclear collision
and the hard scattering, respectively. The variable $\hat q_0$ here 
represents the broadening rate corresponding to quark production at impact 
distance $\tau=0$ in a central collision with $b=0$ at the moment of time $t_0=l_0$. 
The transport coefficient for a gluon is 
by a factor of $9/4$ larger due to the Casimir factor $C_A/C_F$, where
for $N_c = 3$ the factors $C_A = N_c = 3$ and $C_F = (N_c^2-1)/2\,N_c = 4/3$ 
are the strength of the gluon self-coupling and a gluon coupling to a quark, respectively.
The results are practically not sensitive to model-dependent $t_0$, and we fix it naively 
at $t_0=1.0\fm$ which is sufficient for our purposes here. Then, high-$p_T$ hadron production 
is considered as a probe for the medium properties via the single fitted parameter 
$\hat q_0$ in Eq.~(\ref{340}), which depends on energy and atomic number 
$A$ of the colliding nuclei.

The dipole in initial state has a small separation
$r\sim 1/k_T$. During its propagation through the medium, it expands 
with a rate deretmined by the corresponding uncertainty princpile
$dr/dt = k_T(t) [1/\alpha p_T + 1/(1-\alpha) p_T]\propto1/r(t)$ 
(for more details see Refs.~\cite{psi,psi-bnl,green,my-alice}). 
Then, the function $r(l)$ is given by
%
\beq
\frac{dr} {dl}
=\frac{1} {r(l) E_h \alpha (1 - \alpha)}\,.
\label{360}
\eeq
%
Here, the quark momentum fraction in the dipole is $\alpha$, and $E_h = 
p_T$ is the dipole (hadron) energy. Solution of equation (\ref{360}) is given by
%
\beq
r^2 (l) = \frac{2l} {\alpha (1 - \alpha) 
p_T } + r_0^2\,,
\label{380}
\eeq
which is then used in (\ref{302}) for analysis of medium attenuation of the dipole 
determined by $\hat q$,
%
\beq
R_{AB}(\vec{b},\vec\tau,p_T) =
\int\limits_0^{2\pi}
\frac{d\phi}{2\pi} \exp\Biggl[-\frac{4}{p_T}\!
\int\limits_L^\infty \!\!dl \,l\,\hat q(l,\vec{b},\vec\tau+\vec l)
\Biggr]\,.
\label{400}
\eeq
%
This expression represents the attenuation factor for a dipole propagating in a dense medium.
This dipole has been produced by a high-$k_T$ parton (with $k_T = p_T/z_h$) 
at impact distance $\vec\tau$ in a heavy nuclei $A$ and $B$ collision happened at impact distance 
$\vec b$. The lower intergration limit is given by $L=\max\{ l_p, l_0 \}$, while 
the transport coefficient is found from Eq.~(\ref{340}). Finally, in this calculation
we employ the Berger's approximation \cite{berger} and take a symmetric dipole
configuration with $\alpha \simeq 1/2$ where the projected hadron wave function
is at maximal value \cite{radyushkin}.

Integrating Eq.~(\ref{400}) over the impact parameter $\vec{\tau}$
of the hard collision one obtains the nuclear modification factor for inclusive
high-$p_T$ hadron production in heavy ion $A + B$ collision
at relative impact parameter $b$,
%
\beqn
R_{AB}(\vec{b},p_T) =
\frac{\int d^2\tau \,T_A(\tau)T_B(\vec b-\vec\tau)
R_{AB}(\vec{b},\vec\tau,p_T)}
{\int d^2\tau \,T_A(\tau)T_B(\vec b-\vec\tau)}\,.
\label{540}
\eeqn
%

This simplified approach was first employed in Ref.~\cite{my-alice} and provides a rather good 
description of the first data from the ALICE experiment \cite{alice-data} for central 
collisions assuming the gluon jets contribution to leading hadron production
only. Eqs.~(\ref{400}) and (\ref{540}) generalize this simplified model making 
it suitable also for the analysis of suppression at different centralities.
Such an analysis at smaller RHIC energies requires an additional contribution 
of quark jets to leading hadron production. Consequently, the mean production 
length $\la l_p\ra$ is different for quark and gluon jets as was demonstrated 
in Ref.~\cite{green}. Therefore, the numerator in Eq.~(\ref{540}) is calculated 
separately for quark and gluon jets whose contributions are then summed up 
with the weights given by Eq.~(\ref{300}).

Although this simplified heuristic model allows to understand the main features 
of the underlying dynamics, Eq.~(\ref{360}) does not describe 
the expansion of the dipole in a medium where the color filtering effects 
modify the path/length dependence of the mean dipole separation.
Namely, dipoles of large size are strongly absorbed, while small
dipoles attenuate less. Correspondingly, the mean transverse separation 
in a dipole propagating in a medium should be smaller than in vacuum. 
Introducing an absorptive term we arrive at a modified evolution equation,
%
\beq
\frac{dr^2}{dl}=\frac{2}{E_h\alpha(1-\alpha)} - \frac{1}{2} r^4(l)\,\hat q(l),
\label{410}
\eeq
%
which is solved analytically with respect to an explicit form 
for the mean dipole size squared,
%
\begin{widetext}
\beqn
\hspace*{-4.0cm}
r^2(l) = 2\,X\,l\,\Theta(l_0-l)~ - ~
\Theta(l - l_0)\,
\frac
{\sqrt{Z l}}{Y}
\times~~~~~~~~~~~~~~~~~~~~~~~~~~~~~~~~~~~~~~~
\nonumber\\
\frac{
I_1(\sqrt{Z l})
\biggl
( Z\, l_0\,                 K_0(\sqrt{Z l_0  })/2 +
 \sqrt{Z\, l_0}\,           K_1(\sqrt{Z l_0  }) 
\biggr
) 
+
\biggl
( Z\, l_0\,                 I_0(\sqrt{Z l_0  })/2 -
 \sqrt{Z\, l_0}\,           I_1(\sqrt{Z l_0  }) 
\biggr
)
                          K_1(\sqrt{Z l})
}
{
I_0(\sqrt{Z l})
\biggl
( Z\, l_0\,                 K_0(\sqrt{Z l_0  })/2 +
\sqrt{Z\, l_0}\,            K_1(\sqrt{Z l_0  }) 
\biggr
)
+
\biggl
( Z\, l_0\,                 I_0(\sqrt{Z l_0  })/2 -
\sqrt{Z\, l_0}\,            I_1(\sqrt{Z l_0  }) 
\biggr
) 
                          K_0(\sqrt{Z l})
}\,,
\label{420}
\eeqn
\end{widetext}
%
where we neglect the initial dipole size $r_0$, $\Theta(x)$
represents the step function; $I_0(x)$, $I_1(x)$,
$K_0(x)$ and $K_1(x)$ are the modified Bessel functions and factors
$X = 1/\alpha(1-\alpha) p_T$, $Y = \hat{q}(l) l$ with $Z = 4 X Y$.
%
%
%
\begin{figure}[bth]
   \includegraphics[height=13cm,width=8.5cm]{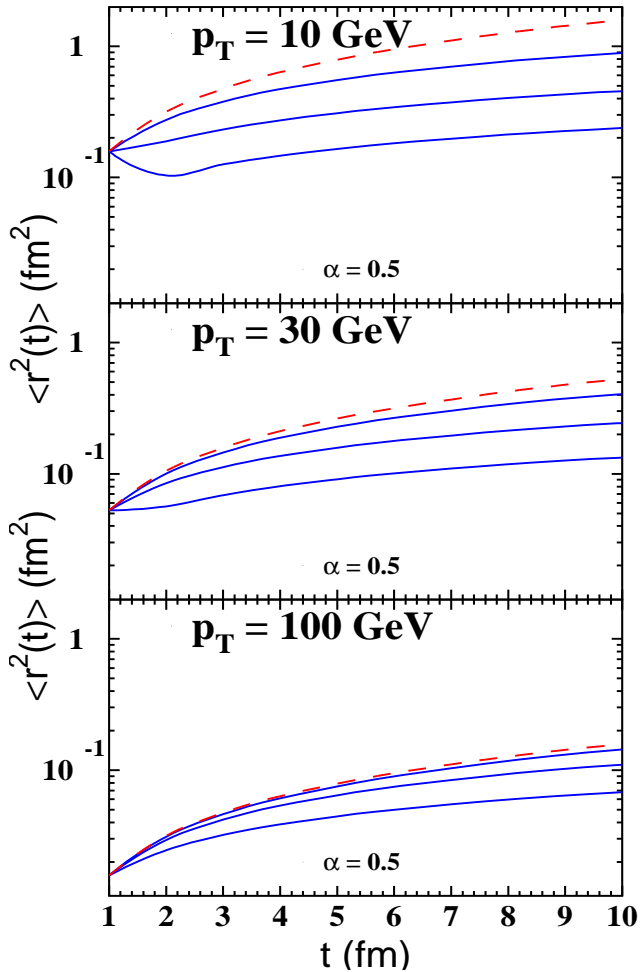}
   \caption{ \label{fig:r2t}(color online) 
           Time evolution of the mean dipole size squared at $p_T = 10$, 30 and
           $100\,\GeV$. Dashed and solid curves are computed within the 
           simplified model without (Eq.~(\ref{380})) and with (Eq.~(\ref{410}))
           color filtering effects, respectively. Solid curves
           are computed for different fixed values of $\hat{q}_0$ = 0.1,
           0.5 and 2.0 $\GeV^2/\fm$ from top to bottom.
           } 
\end{figure}
%

In comparison with Eq.~(\ref{360}), the color filtering effects in Eq.~(\ref{410}) result 
in a reduction of the mean dipole size demonstrated in
Fig.~\ref{fig:r2t} for fixed values of $p_T$ = 10, 30 and 100 $\GeV$. 
Here, we fix $\alpha=1/2$ again \cite{berger}. Such a reduction 
makes the medium more transparent and depends strongly on $\hat{q}_0$. 
The larger is $\hat{q}_0$, the stronger is the reduction. Fig.~\ref{fig:r2t} also 
demonstrates that the reduction gradually decreases with an increase of $p_T$. 
Correspondingly, we should expect that the analysis of ALICE data performed 
in Ref.~\cite{my-alice} should lead to an underestimated medium density, i.e. 
to a smaller parameter $\hat q_0$. 

As was mentioned in Sect.~\ref{hadronization}, during the short length
between $l_0$ and $l=l_p>l_0$ in the medium the parton 
undergoes multiple interactions. The latter trigger an additional gluon 
radiation and consequently an extra energy loss \cite{bdmps}.
Corresponding expression for energy loss comes from Eq.~(\ref{100}) and reads
%
\beq
\Delta E=\frac{3\alpha_s}{4}\,
\Theta(l_p-l_0)
\int\limits_{l_0}^{l_p} dl
\int\limits_{ l_0 }^l 
dl'\, \hat q( l' )\,.
\label{550}
\eeq
%

Although this is a small correction, we have explicitly included it in the calculations 
by making a proper shift of the variable $z_h$ in the fragmentation function.

\section{Numerical results vs data}
\label{data}
%

Here, we compare results of our simple model including the color filtering effects with 
numerous data available from the recent precision measurements at RHIC and LHC.

%
\subsection{Hadron quenching at high-\boldmath$p_T$}
\label{RAA}
%

Comparison of the $R_{AA}(b=0,p_T)$ factor calculated above in the framework 
of considered model with the first data from the ALICE experiment \cite{alice-data} at 
$\sqrt{s}=2.76\TeV$ was performed in Ref.~\cite{my-alice}. The maximal 
value of the transport coefficient was adjusted to these data
and fixed at $\hat q_0=0.4\GeV^2\!/\!\fm$. The growing $p_T$-dependence 
of $R_{AA}(p_T)$ follows from the reduction of the mean dipole size 
at higher $p_T$ (in accordance with Eq.~(\ref{380})) and 
due to a Lorentz dilation of the dipole size expansion. 
This leads to a more transparent medium for more energetic smaller dipoles in 
accordance with the CT effect. An analogous growing energy dependence of 
the medium transparency was predicted and observed in the case of virtual 
photoproduction of vector mesons on nuclei \cite{CT}.
%
%
\begin{figure}[bht]
   \vspace*{5mm}
   \centerline{
         \scalebox{0.45}{\includegraphics{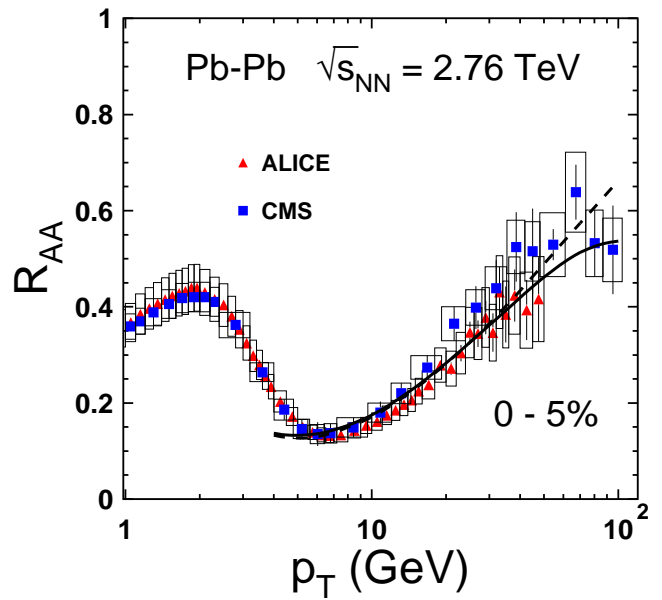}}}
   \caption{\label{fig:lhc-0-5} (color online) The nuclear modification factor 
           $R_{AA}(p_T)$ for charged hadron production in central 
           (with centrality $0-5\%$) Pb-Pb collisions at 
           $\sqrt{s}=2.76\TeV$. The dashed line represents 
           the simple model prediction given by Eqs.~(\ref{302}), 
           (\ref{420}), (\ref{540}), and the space- and 
           time-dependent transport coefficient (\ref{340}) with the 
           single adjustable parameter $\hat q_0=1.3\GeV^2\!/\!\fm$.
           The solid curve, in addition, includes the effect of initial 
           state interactions in nuclear collisions \cite{isi,kn-review} 
           described in Sect.~\ref{isi}. The data for $R_{AA}(p_T)$ 
           are from the ALICE \cite{alice-new} and CMS 
           \cite{cms-new1,cms-new2} measurements.}
\end{figure}
%
%
%

Being encouraged by the first success of the simplified model, we have generalized 
this model for studies of production of various high-$p_T$ hadrons at distinct energies 
by incorporating additional quark jet contributions to leading hadron production besides 
gluon ones with relative weights given by Eq.~(\ref{300}). The original treatment, which was 
suitable only for central heavy ion collisions, has now been extended also to non-central 
collisions. Moreover, for the first time we have incorporated also the color filtering effects in the 
evolution equation for the mean dipole size, Eq.~(\ref{410}).

The results for central ($0-5\%$) Pb-Pb collisions at $\sqrt{s}=2.76\TeV$ 
are shown by dashed curve in Fig.~\ref{fig:lhc-0-5}, 
compared to the ALICE data \cite{alice-new} and CMS 
\cite{cms-new1,cms-new2}. The only free parameter, the maximal 
value of the transport coefficient defined in Eq.~(\ref{340}),
was adjusted to the data and fixed at $\hat q_0=1.3\GeV^2\!/\!\fm$ 
corresponding to the mean value of the impact parameter
$\la b\ra = 2.3\,\fm$. This value of the transport
coefficient is then valid for all further calculations 
for Pb-Pb collisions at c.m. energy $\sqrt{s} = 2.76\,\GeV$.
%
%
\begin{figure}[htb]
   \vspace*{5mm}
   \centerline{
         \scalebox{0.45}{\includegraphics{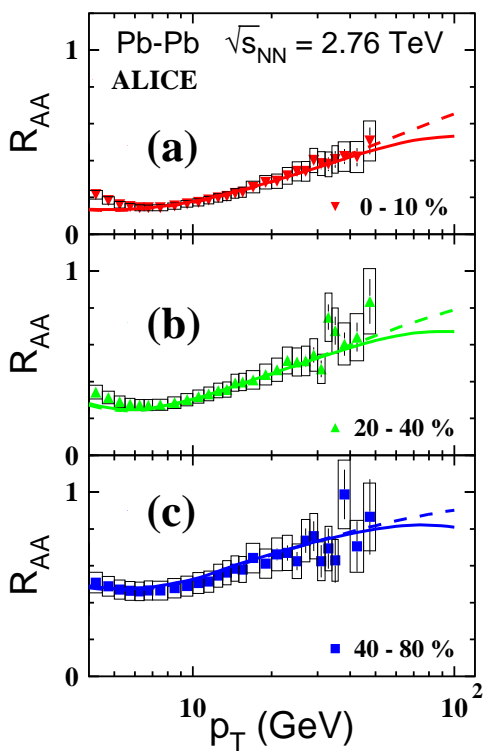}}}
   \caption{\label{fig:alice-b} (color online) The centrality 
    dependence of the nuclear modification factor $R_{AA}(p_T,b)$ 
    for charged hadron production
    measured by the ALICE experiment \cite{alice-new}. 
    The intervals of centrality are indicated. 
    The meaning of the curves is the same as in Fig.~\ref{fig:lhc-0-5}. }
 \end{figure}
%

Note, while our calculations describe well the data at high $p_T\gsim 
6\GeV$, the region of smaller $p_T$ is apparently dominated by thermal 
mechanisms of hadron production.

\begin{figure}[tbh]
   \vspace*{5mm}
   \centerline{
         \scalebox{0.45}{\includegraphics{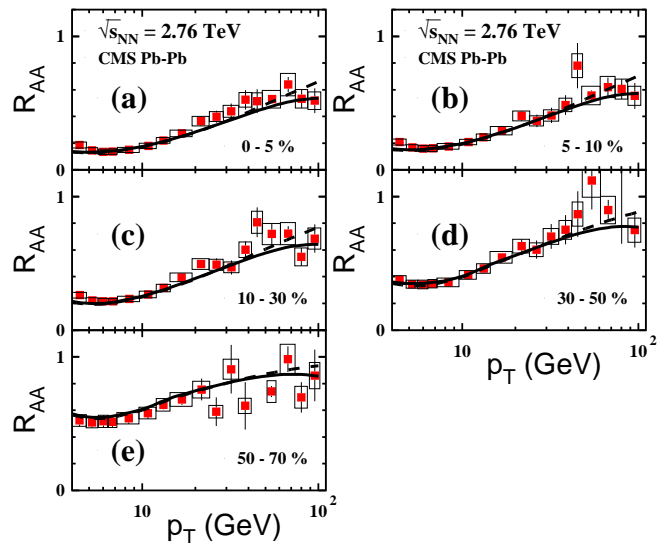}}}
   \caption{\label{fig:cms-b} (color online) The same as in 
           Fig.~\ref{fig:alice-b}, but with the data from CMS 
           \cite{cms-new1,cms-new2}. }
\end{figure}

The variation of the suppression factor $R_{AA}(b,p_T)$ with impact parameter 
$b$ given by Eqs.~(\ref{302}), (\ref{420}) and (\ref{540})
is plotted by dashed curves and compared to the data at different centralities 
obtained by ALICE \cite{alice-new} in Fig.~\ref{fig:alice-b}, 
and by CMS \cite{cms-new1,cms-new2} in Fig.~\ref{fig:cms-b}. 
In all the cases we observe a rather good agreement.
%
%

%
\subsection{Large \boldmath$x_T$ suppression due to energy conservation}
\label{isi}
%

In the course of propagation through the nucleus multiple interactions 
of the projectile hadron and its debris lead to a dissipation of energy. 
The corresponding loss of energy is proportional to the energy 
of the projectile hadron, thus the related effects do not 
disappear at very high energies as was stressed in Ref.~\cite{isi} 
(see also Ref.~\cite{peigne}). In the Fock state representation, 
the projectile hadron can be decomposed over different states, which
are the fluctuations of this hadron ``frozen" by Lorentz time dilation.  
The interaction with the target modifies the weights of Fock states since
some interact stronger, some weaker. 

In each Fock component the hadron momentum is shared between its
constituents, and the momentum distribution depends on their multiplicity: 
the more constituents are involved, the smaller is the
mean energy per parton. This leads to the softer
fractional energy distribution of a leading parton, and
the projectile parton distribution falls at $x\rightarrow 1$ steeper
on a nuclear target than on a proton. 

In the case of hard reaction on a nucleus, such softening 
of the projectile parton fractional energy distribution 
can be viewed as an effective energy loss of the 
leading parton due to initial state multiple interactions.
This leads to an enhancement of the weight factors
for higher Fock states in the projectile hadron with a large
number of constituents having a tough energy sharing.
Consequently, the mean energy of the leading parton decreases
compared to lower Fock states which dominate the hard
reaction on a proton target. Such a reduction of the mean fractional 
energy of the leading parton is apparently independent of the initial
hadron energy and can be treated as an effective loss of energy 
proportional to the initial hadron energy. A detailed description 
and interpretation of the corresponding additional suppression 
was presented also in Refs.~\cite{isi-jan,green,conservation14}.

Here we would like to emphasize that this effective energy loss
is different from the energy loss by a single parton propagating
through a medium and experiencing induced gluon radiation.
In this case the mean fractional energy carried by the radiated 
gluons vanishes at large initial energies $E$
as $\Delta E/E\propto1/E$ \cite{feri,bh,bdmps}.

The initial state energy loss (ISI effect) is a minor effect 
at high energies and midrapidities. However, it may 
essentially suppress the cross section 
when reaching the kinematical bounds,
$x_L = 2p_L / \sqrt{s} \to 1$ and $x_T = 2p_T / \sqrt{s} \to 1$. 
Correspondingly, the proper variable which controls this 
effect is $\xi=\sqrt{x_L^2+x_T^2}$. 

The magnitude of suppression was evaluated in 
Ref.~\cite{isi,kn-review}. It was found 
within the Glauber approximation that 
each interaction in the nucleus
leads to a suppression $S(\xi)\approx 1-\xi$.
Summing up over the multiple initial state interactions
in a p-A collision with impact parameter $b$ and
relying on  Eq.~(\ref{300}), 
one arrives at a nuclear ISI-modified PDF 
$F_{i/p} (x_i, Q^2) \Rightarrow 
F^{(A)}_{i/p} (x_i, Q^2, b)$,
where
%
 \beqn
\hspace*{-0.40cm}
F^{(A)}_{i/p}
(x_i,Q^2,b)&=&
C\,F_{i/p}(x_i,Q^2)\,
\nonumber\\ 
&\times&
\frac{
\left[
e^{-\xi\sigma_{\rm eff}T_A(b)}-
e^{-\sigma_{\rm eff}T_A(b)}\right]}
{(1-\xi)\,\left[1-
e^{-\sigma_{\rm eff} T_A(b) } \right]}\,,
\label{600}
 \eeqn
%
where $\sigma_{\rm eff} = 20\mb$ \cite{isi,lrg} is the hadronic cross section which
effectively determines the rate of multiple interactions, and constant $C$ 
is given by the Gottfried sum rule. It was found that such an additional nuclear suppression 
due to the ISI effects represents an energy independent feature common 
for all known reactions, experimentally studied so far, with any leading particle 
(hadrons, Drell-Yan dileptons, charmonium, etc.). Following to Ref.~\cite{kn-review}, 
in the analysis of high-$p_T$ hadron production we apply exactly the same model 
developed in Refs.~\cite{isi,isi-jan,conservation14} for large $x_L$ and 
with the same physical parameters.

Using PDFs modified by the ISI energy loss, Eq.~(\ref{600}),
one achieved a good parameter-free description of available data from the BRAHMAS 
\cite{brahms} and STAR
\cite{star-forward} experiments at forward rapidities (large $x_L$) in $dA$ collisions
\cite{isi,kn-review}. An alternative interpretation \cite{kkt-04} at forward rapidities
is based on the coherence effect (color glass condensate or CGC), which should
disappear at lower energies because $x\propto 1/\sqrt{s}$ increases. 
Assuming that besides CGC there is no other mechanism contributing to the suppression 
observed by BRAHMS, one should not expect any suppression at smaller energies
where no coherence effects are possible. However, according to Eq.~(\ref{600})
the suppression caused by the ISI energy loss scales in Feynman 
$x_F = x_L$ and should exist at any energy. Thus, by reducing the collision energy
one should provide a sensitive test for the models. Expectation of no suppression 
following from CGC at forward rapidities and small energies is in contradiction 
with data from the NA49 experiments \cite{na49} at SPS obtained at much smaller 
energy than BRAHMS. This observation confirms that the effect of suppression 
at forward rapidities is still there and can be explained entirely by the ISI energy loss. 

\begin{figure}[tbh]
   \vspace*{-9mm}
   \centerline{
         \scalebox{0.45}{\includegraphics{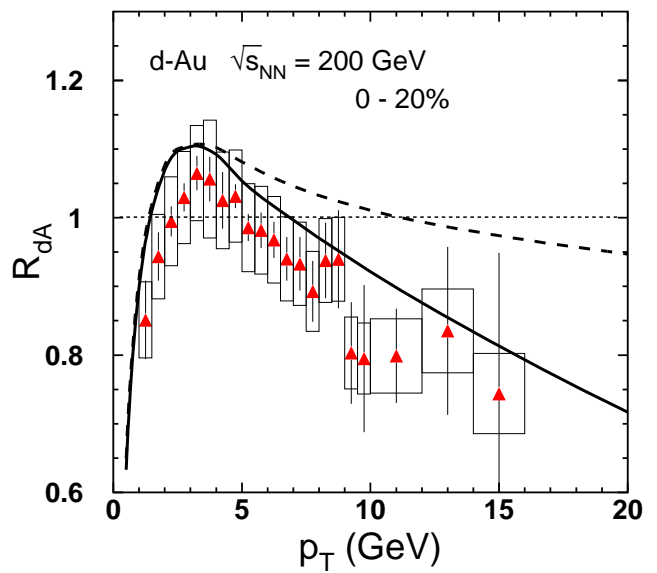}}}
   \caption{\label{fig:cronin} 
    (color online)
         The nuclear modification factor $R_{dAu}(p_T)$ for  
         neutral pions produced in central
         ($0-20\%$) d-Au collisions
         at $\sqrt{s}=200\GeV$ and $\eta = 0$. 
         The solid and dashed curves show the predictions 
         calculated with and without the ISI corrections, respectively. 
         The isotopic effect is included. The data are 
         from the PHENIX experiment \cite{cronin-phenix}.}
\end{figure}

Another test of the mechanism based on the ISI energy loss can be realized for RHIC
energies at large $p_T$, where no coherence effects are expected, while the value of 
$\xi$ is considerable and an ISI energy deficit should lead to a suppression at
large $x_T$ similar to what was observed at large $x_L$. The predictions 
for large-$p_T$ suppression based on ISI effects \cite{kn-review} at the 
collision energy $\sqrt{s} = 200\,\GeV$ and at midrapidity are confirmed 
by the PHENIX data for central d-Au collisions \cite{cronin-phenix} as is demonstrated in
Fig.~\ref{fig:cronin}. No alternative explanation has been proposed so far.
%
%

The ISI energy loss also affects the $p_T$ dependence of the nuclear suppression in
heavy ion collisions. These effects are calculated similarly to p(d)-A collisions using 
the modified parton distribution functions Eq.~(\ref{600}) for nucleons in both colliding 
nuclei. Naively, at the LHC one would not expect any sizable ISI energy loss. 
Nevertheless, even at such high energies one can reach large $x_T$ sufficient for 
the manifestation of ISI energy loss effects. These corrections are predicted to be 
important for $p_T\gsim 70\,\GeV$. This is demonstrated
in Figs.~\ref{fig:lhc-0-5}, \ref{fig:alice-b} and \ref{fig:cms-b} by solid lines
which show a flattening of the $R_{AA}(p_T)$ factor at large $p_T$.

\begin{figure}[tbh]
   \vspace*{5mm}
   \centerline{
         \scalebox{0.45}{\includegraphics{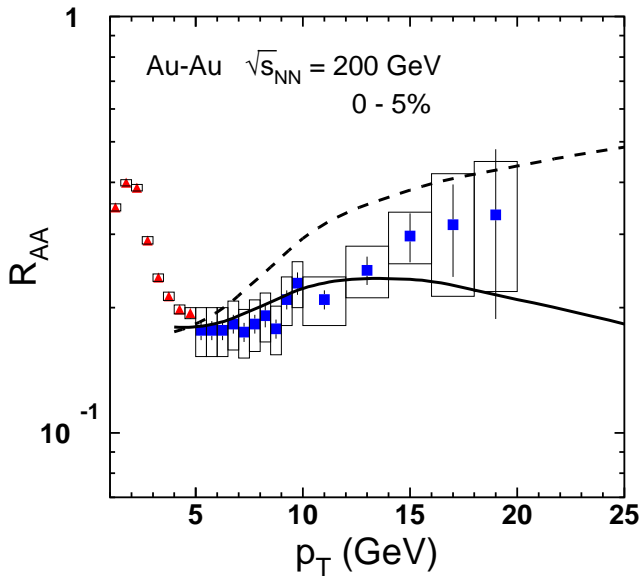}}}
   \caption{\label{fig:rhic-0-5} (color online) The nuclear modification factor 
           $R_{AA}(p_T)$ for neutral pions produced in central Au-Au 
           collisions at $\sqrt{s}$ = 200 $\GeV$.
           The solid and dashed line are computed with or without
           ISI corrections, respectively.
           The PHENIX data are from Refs.~\cite{phenix-b} (triangles)
           and \cite{phenix-0} (squares).}
\end{figure}

At RHIC energies this additional suppression reduces $R_{AA}(p_T)$ 
significantly at large $p_T$ as is demonstrated in Fig.~\ref{fig:rhic-0-5}
for central Au-Au collisions at $\sqrt{s}=200\GeV$.
Including the CT effect only and repeating the same calculation as was done 
above for the LHC, we get $R_{AA}$ steeply rising with $p_T$, 
as is depicted by the dashed curve. Since parameter $\hat{q}_0$ is expected 
to vary with energy, it was re-adjusted and found to be $\hat{q}_0(RHIC)=1.0\GeV^2\!/\!\fm$, 
which is less than in collisions at the LHC\footnote{Notice that $\hat{q}_0$ is also $A$-dependent.}.
Inclusion of the ISI effects, Eq.~(\ref{600}), leads to a sizeable additional suppression, as is shown 
by the solid curve.
%
%
%
%
\begin{figure}[tbh]
  \centerline{
   \scalebox{0.45}{\includegraphics{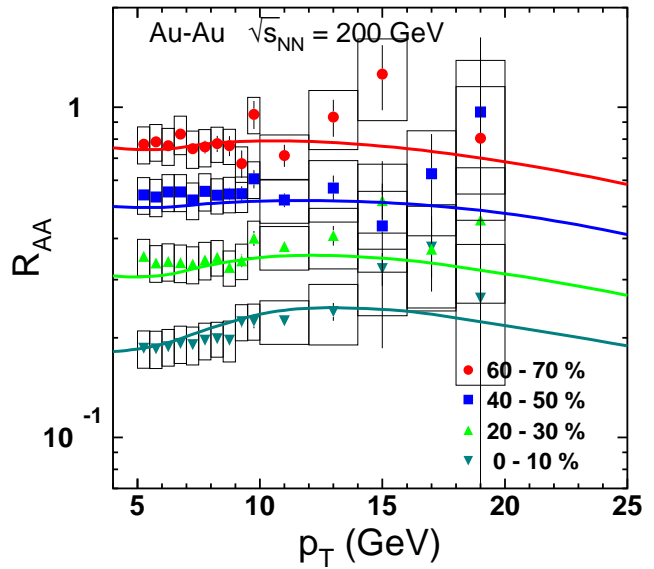}}}
   \caption{\label{fig:rhic200-b-gf} (color online) 
   The centrality dependence of the nuclear modification 
   factor $R_{AA}(p_T,b)$ in comparison to the PHENIX data 
   with Au-Au collisions at $\sqrt{s}=200\GeV$ \cite{phenix-b}.}
\end{figure}
%
%

By fixing $\hat q_0 = 1.0\,\GeV^2/\fm$ we can obtain other observables for
Au-Au collisions at $\sqrt{s}=200\GeV$. Fig.~\ref{fig:rhic200-b-gf} shows 
our results for the suppression of $\pi^0$ at different centralities, 
in comparison with the PHENIX data \cite{phenix-b}.

\begin{figure}[tbh]
   \vspace*{0mm}
   \centerline{
         \scalebox{0.45}{\includegraphics{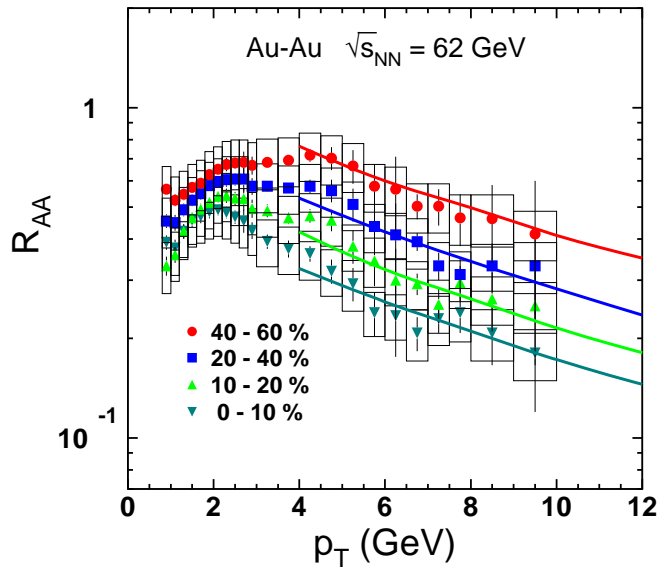}}}
   \caption{\label{fig:rhic62-b} (color online) The same as in 
           Fig.~\ref{fig:rhic200-b-gf}, but at $\sqrt{s}=62\GeV$. 
           The data are taken from Ref.~\cite{phenix62-b}.}
\end{figure}

One achieves stronger ISI effects either by enlarging 
$p_T$ at some fixed energy, or by reducing the collision energy
keeping the $p_T$ range unchanged. The latter is demonstrated by
the data on $R_{AA}(p_T)$ in Au-Au collisions at smaller RHIC energy 
$\sqrt{s}=62\GeV$ in Fig.~\ref{fig:rhic62-b} showing an 
increasing rather than falling nuclear suppression as a direct
manifestation of the ISI energy loss effects, Eq.~(\ref{600}). 
Such data are important for studies of the energy loss effects.
Our calculations again demonstrate a good agreement with the data. 
Here, the hot medium properties are changed, so in order to account 
for that we have re-adjusted the parameter $\hat q_0=0.7\GeV^2\!/\!\fm$.
%
%
%
%

%
\subsection{Anisotropy in azimuthal angle distribution}
\label{v2}
%

The observed high-$p_T$ hadrons suppression reflects 
a contribution of some effective medium volume.
Thus one should expect that the resulting suppression depends on 
the propagation length while the direction of propagation perpendicular
to the surface of the medium is preferrable. This effect leads to 
an azimuthal asymmetry in the hadron angular distribution presumably
for non-central collisions \cite{green}.

\begin{figure}[tbh]
   \centerline{
         \scalebox{0.45}{\includegraphics{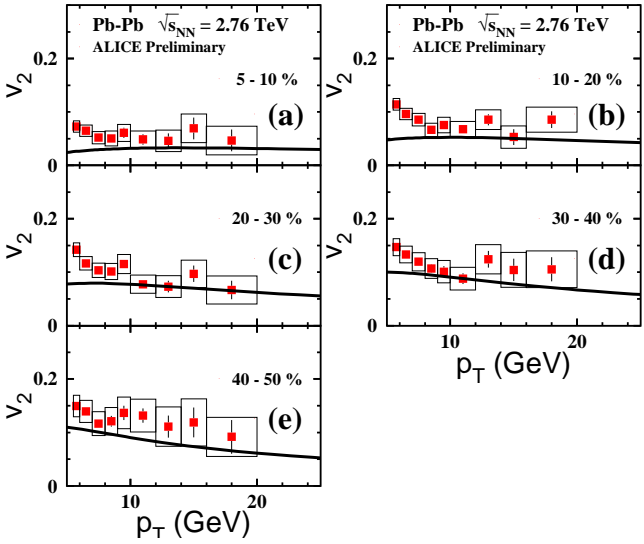}}}
   \caption{\label{fig:alice-v2-all} (color online) The azimuthal 
   anisotropy, $v_2$, vs $p_T$ in comparison to ALICE data 
   \cite{alice-phi-v2} for charge hadron production in Pb-Pb 
   collisions at midrapidity, at $\sqrt{s}$ = 2.76 $\TeV$ 
   and at different centralities.}
 \end{figure}

\begin{figure}[tbh]
   \vspace*{5mm}
   \centerline{
         \scalebox{0.45}{\includegraphics{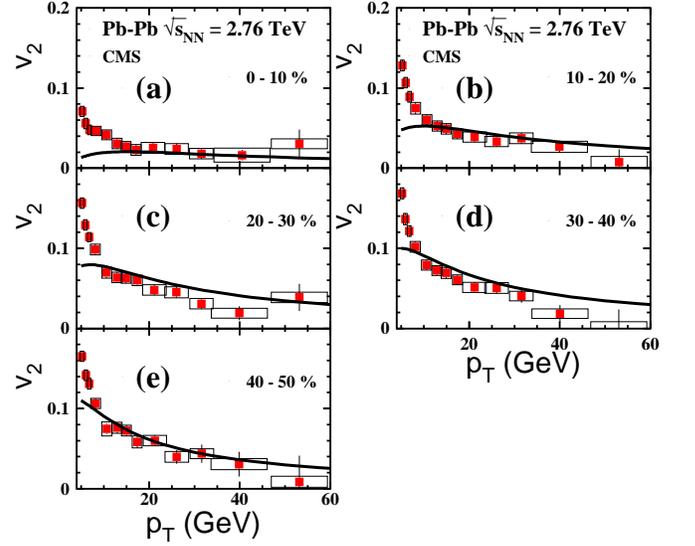}}}
   \caption{\label{fig:cms-v2-all} (color online) The same as in 
           Fig.~\ref{fig:alice-v2-all}, but with the CMS data \cite{cms-v2}.}
\end{figure}

Typically, for azimuthal asymmetry one studies the second moment of the 
azimuthal angle distribution, i.e. $v_2\equiv \left \la \cos(2 \phi) \right \ra$ given by
%
\beq
  v_2(p_T,b)=
  \frac{\int\hspace*{-0.1cm} d^2\tau 
  T_A (\tau) T_B (\vec b\!-\!\vec\tau)\!\int\limits_0^{2\pi}\!\!d\phi
  \cos(2\phi)
  R_{AB}^{\phi} (\vec{b}, \vec \tau, p_T)
  }
  {\int d^2 \tau T_A( \tau)
  T_B(\vec b-\vec\tau)
  \int\limits_0^{2\pi}d\phi 
  R_{AB}^{\phi}(\vec{b},\vec\tau,p_T)
  }\,,
\label{620}
\eeq
%
where
$R_{AB}^{\phi}(\vec{b},\vec\tau,p_T)\equiv 
 R_{AB}(\vec{b},\vec\tau,p_T,\phi)$ is given by Eq.~(\ref{400})
but without integration over $\phi$. The results of this calculation
are compared with ALICE \cite{alice-phi-v2} and CMS data \cite{cms-v2} in 
Figs.~\ref{fig:alice-v2-all} and \ref{fig:cms-v2-all} and are found to be in
a good overall agreement. 
%
%
%
%
%
%
%
%

Similarly to the data on $R_{AA}$, our pQCD calculations for $v_2(p_T)$ 
grossly underestimate the data at small $p_T\lesssim6\GeV$ due to the
presence of two different mechanisms discussed earlier in Ref.~\cite{green}: 
the dominant hydrodynamic mechanism of elliptic flow, leading to a large 
and increasing anisotropy $v_2(p_T)$ at high $p_T$'s, which abruptly 
switches to the pQCD regime having a much smaller azimuthal anisotropy.

\begin{figure}[h!]
   \vspace*{5mm}
   \centerline{
       \scalebox{0.45}{\includegraphics{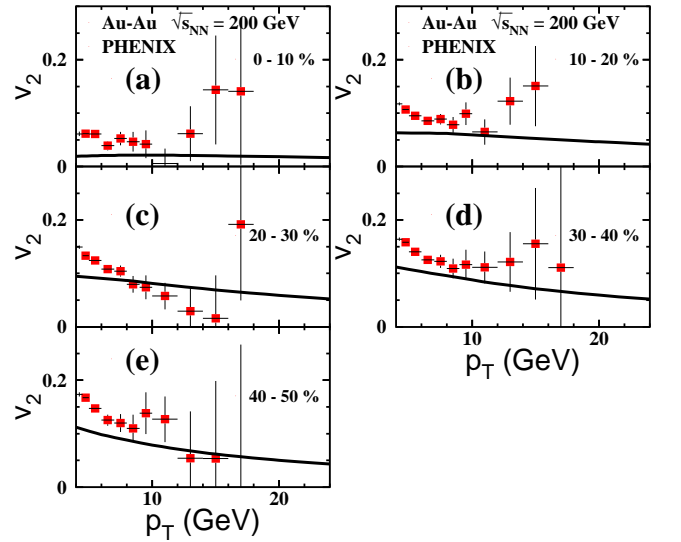}}}
   \caption{\label{fig:rhic-v2-all} (color online) The same as in 
           Fig.~\ref{fig:alice-v2-all}, but with the PHENIX data \cite{phenix-v2}.}
\end{figure}

Finally, we have also evaluated the azimuthal anisotropy at smaller c.m. 
energies and compared with the corresponding RHIC data. Our results
agree well with PHENIX data for $\pi^0$ production which is 
demonstrated in Fig.~\ref{fig:rhic-v2-all}.

%
\section{Summary}
%

In this paper, we contribute to a quantitative 
understanding of the strong nuclear suppression 
of leading high-$p_T$ hadrons produced inclusively 
in heavy ion collisions. The main motivation 
comes from the improved quality of data at RHIC 
and new high-statistics data at LHC which 
provide a strong potential for a more decisive 
verification of different models.
%
%
%
%

The popular pure energy-loss scenario, 
based on the unjustified assumption of long
production length, experiences difficulties 
explaining the data and also fails
to describe simultaneously the data from the HERMES experiment for leading
hadron production in semi-inclusive DIS, 
which represents a sensitive testing
keystone for in-medium hadronization models.

In this work, we present an alternative mechanism 
for nuclear suppression in high-$p_T$ hadron production. 
The key point is that the production length 
of leading hadrons does not rise with $p_T$ 
and remains rather short as is demonstrated in Fig.~\ref{fig:mean-lp}
(see also Refs.~\cite{my-alice,green}). This is a consequence of a strong increase 
of the energy dissipation by a highly virtual parton, produced in a high-$p_T$ process, 
with jet energy. The production moment of a colorless hadronic state 
(``pre-hadron'') ceases the energy loss. The main reason for such a suppression 
is related to the survival probability of the ``pre-hadron'' propagating through 
the dense matter. According to Eqs.~(\ref{380}) and (\ref{420}), 
larger $p_T$'s prefer a smaller dipole size and the medium becomes more transparent 
in accordance with the color transparency effect. 
The corresponding increase of the nuclear suppression factor 
$R_{AA}(p_T)$ is indeed observed at the LHC.

Thus we conclude that the main reason for the observed suppression of high-$p_T$
hadron production in heavy ion collisions is not an energy loss, but attenuation 
of early produced colorless dipoles (``pre-hadrons'') propagating through
a dense absorptive matter. 

One should clearly discriminate between vacuum and medium-induced energy loss. 
The former is much more intensive and is the main cause of shortness of
the production length $l_p$.

One should also discriminate between the terms ``jet quenching'' and ``hadron quenching''. 
The latter is considered in the present paper, namely, a high-$p_T$ hadron detected 
inclusively and carrying the main fraction $z_h$ of the accompanying jet energy as is
depicted in Fig.~\ref{fig:mean-zh}. This leads to a smallness of $l_p$ scale as
a consequence of constraints coming from an intensive energy loss in vacuum
at large $p_T$ and energy conservation. On the other side, if none of hadrons
in the jet are forced to have a large $z_h$ and the whole jet has a large $p_T$, 
then the hadronization lasts a long time due to the usual Lorentz time dilation.

Although the attenuation of colorless dipoles propagating through a medium 
has already been studied in Ref.~\cite{green} within the rigorous 
quantum-mechanical approach based on the Green functions formalism, in this paper we 
present an alternative simplified description of the high-$p_T$ hadron production 
in heavy ion collisions. For this purpose, we start from the simple model 
of Ref.~\cite{my-alice} and generalize it for non-central heavy ion collisions 
additionally including quark jet contributions to leading hadron production. 
In the framework of this generalized model, we have incorporated 
for the first time the color filtering effects 
in the evolution equation for the mean dipole size and found 
its simple analytical solution, Eq.~(\ref{420}). 
As the main result of this study, we found that the color filtering effects lead 
to a reduction of the mean dipole size as demonstrated in Fig.~\ref{fig:r2t} and, 
consequently, to a more transparent nuclear medium. 

First, we have studied the suppression factor $R_{AA}(p_T)$ in comparison to available data 
at various centralities and c.m. energies in corresponding RHIC and LHC 
kinematic domains. At large $x_T\gsim0.1$ we included an additional suppression factor 
from the initial-state interactions. 
This factor, which falls steeply with $p_T$, causes a rather flat
$p_T$ dependence of $R_{AA}(p_T)$ function at RHIC energy $\sqrt{s} = 200\,\GeV$ 
and even leads to an increase of suppression at high $p_T$'s at lower $\sqrt{s} = 62\,\GeV$. 
At the next step, we evaluated the azimuthal anisotropies in hadron production 
at different energies and centralities corresponding to measurements at RHIC and LHC.

In all the cases we found a fair agreement with data at high $p_T$'s. 
The only adjustable parameter which is the maximal magnitude of 
the transport coefficient, was estimated between $\hat q_0=1.3 \GeV^2 \!/ \!\fm$, 
$1.0 \GeV^2 \! / \! \fm$ and $0.7 \GeV^2 \!/ \!\fm$ values
at energies $\sqrt{s}=2.76\TeV$, $200\GeV$ and $62\GeV$, respectively, 
for an initial time scale $t_0 = 1\fm$ and for such nuclei as Pb and Au. 
These values of $\hat q_0$ are larger than was found 
in the analysis of the first ALICE data \cite{my-alice} 
due to the color filtering effects included in Eq.~(\ref{410}). 
Simultaneously, they are close to the expected magnitude 
$\hat{q}_0\sim 1\,\GeV^2/\fm$ \cite{bdmps} as well as 
to values extracted from RHIC and LHC data by the
JET Collaboration \cite{jet}. Moreover, values of $\hat q_0$ extracted 
within the simplified model are similar and 
only slightly smaller than those found in Ref.~\cite{green} 
within a rigorous quantum-mechanical description.
It is worth noticing that the pure energy loss scenario leads to the value 
of the transport coefficient \cite{phenix-theor}, which is an order 
of magnitude larger than expected \cite{bdmps}.

Remarkably, such a simple formulation absorbs all the important physical effects which 
are naturally inherited from the much more involved Green functions formalism 
and enables immediate comparison of its predictions 
to various existing data on high-$p_T$ hadrons production in heavy ions collisions. 
Thus, technically our model is much more convenient for the practical use 
for testing of the underlined nuclear effects than far more complicated 
path integral techniques.\\

%
%

{\bf Acknowledgments} We are indebted to Boris Kopeliovich for inspiring discussions and useful 
correspondence. This work was supported in part by Fondecyt (Chile) grants 1130543, 
1130549 and by Conicyt-DFG grant No. RE 3513/1-1. The work of J. N. was partially 
supported by the grant 13-20841S of the Czech Science Foundation 
(GA\v{C}R), by the Grant MSMT LG13031, by the Slovak Research and 
Development Agency APVV-0050-11 and by the Slovak Funding Agency, Grant 2/0020/14.

%

\end{document}